\documentclass[12pt]{article}

\usepackage[dvips]{color}                 
\usepackage{graphicx}                     
\usepackage{amssymb}
\usepackage{amsmath}
\usepackage{xspace}                       
\usepackage{rotating}
\usepackage{epsfig}
\newcommand{\beq}{\begin{equation}}
\newcommand{\eeq}{\end{equation}}
\newcommand{\beqa}{\begin{eqnarray}}
\newcommand{\eeqa}{\end{eqnarray}}

\begin{document}

\newcommand{\mytitle}[1]{\begin{center} \LARGE{\textbf{#1}} \end{center}}
\newcommand{\myauthor}[1]{\textbf{#1}}
\newcommand{\myaddress}[1]{\textit{#1}}
\newcommand{\mypreprint}[1]{\begin{flushright} #1 \end{flushright}}

\begin{titlepage}
	\mypreprint{{MIT-CTP-3941} \par}
	\vspace*{1.0cm}
	\mytitle{Chiral symmetry and the axial\\ nucleon to $\Delta\,(1232)$ transition form factors}
	\vspace*{1cm}
	\begin{center}
		\myauthor{Massimiliano Procura} \par
		\vspace*{0.5cm}
		\myaddress{Center for Theoretical Physics, Laboratory for Nuclear Science,\\
		 Massachusetts Institute of Technology, Cambridge, MA 02139, USA\\
			}
		\vspace*{0.2cm}
		\end{center}
	\vspace*{2.5cm}

\begin{abstract}
We study the momentum and the quark mass dependence of the axial nucleon to $\Delta\,(1232)$ transition form factors in the framework of non-relativistic chiral effective field theory to leading-one-loop order. The outcome of our analysis provides a theoretical guidance for chiral extrapolations of lattice QCD results with dynamical fermions.
\end{abstract}

\end{titlepage}
	
\setcounter{page}{2}

\newpage

\sloppy
 
\section{Introduction}

In the past decade, lattice QCD has been developing as a major theoretical tool to quantitatively investigate nucleon structure. Isovector vector and axial form factors \cite{QCDSFconf1,alexNff,yam}, lowest moments of (generalized) parton distributions \cite{gpds,gpdQCDSF}, electromagnetic and axial nucleon to $\Delta\,(1232)$ transition form factors \cite{alexem,al1,alex} are by now calculated using dynamical fermions at pion masses as low as about $350\,{\rm MeV}$ (see also \cite{phil}).
 Chiral effective field theory complements these results by providing a systematic framework to extrapolate to the small quark masses relevant for comparison with phenomenology. 
 As a confluence of recent developments, successful chiral extrapolations in the two-flavor sector have been performed for several nucleon properties, with low-energy parameters in agreement with available information from hadronic processes. Examples of those studies concern the nucleon mass \cite{mass1,mass2,BHMcutoff,BHMdelta,QCDSFmass,QCDSFmass2}, the axial-vector coupling  $g_A$ \cite{ga1,ga2,LHPCga,QCDSFga}, the isovector anomalous magnetic moment \cite{alexem,HW} and the nucleon generalized form factors \cite{gpds,marina}.
  
Here we focus on the axial nucleon ($N$) to $\Delta\,(1232)$ transition, which is relevant for processes of weak pion production off the nucleon. The literature of model calculations in this context is extensive (see {\it e.g.} \cite{ls,svh}), starting from the 1960's: several approaches have been pursued including, for example, the isobar model ({\it e.g.} \cite{bermanvel,liu}), dispersion relations ({\it e.g.} \cite{adler}), non-relativistic and relativistic quark models ({\it e.g.} \cite{rav,lmz,hemmert,barquilla}). Empirical parameterizations of the squared momentum transfer ($q^2$) dependence of the axial $N$ to $\Delta\,(1232)$ form factors have been used to describe the ANL \cite{bar,rad} and BNL \cite{kita1} bubble chamber cross section data for neutrino-induced $\Delta$-resonance production (see also \cite{fn,sehgal,sato,hernandez,alvarez,paschos}). Theoretical input from QCD on this axial transition is important and timely in relation to both current and future neutrino experiments (see {\it e.g.} \cite{k2k,boone,mahn,mini,minerva}) and to the study of parity-violating electroweak excitation of the $\Delta\,(1232)$ with polarized electron scattering at Jefferson Laboratory (see also \cite{muk,zhu2}). The form factors parameterizing the nucleon to $\Delta\,(1232)$ matrix element of the isovector axial quark current have been recently evaluated in lattice QCD \cite{al1,alex} and a revised analysis will be available soon \cite{forth}. 
In this paper we present an analytic calculation which aims at providing a consistent theoretical guidance for chiral extrapolations of the axial $N$ to $\Delta\,(1232)$ transition form factors. A detailed analysis of the new lattice data will be presented in a companion article \cite{forth} in collaboration with the authors of Ref. \cite{alex}. 

 We study the dependence on small $q^2$ and small quark masses of the axial $N$ to $\Delta\,(1232)$ form factors in the framework of the $SU(2)$ non-relativistic chiral effective field theory with pion, nucleon and $\Delta\,(1232)$ degrees of freedom, which is referred to as the `small scale expansion' (SSE) \cite{HHK}. In this scheme, a systematic power counting is established in the small scale $\epsilon$ which denotes, collectively, soft momenta, pion mass and delta-nucleon mass splitting. A leading-one-loop analysis of the $q^2$ dependence of the same form factors, performed in a different framework, has recently appeared \cite{jorge}. This work uses covariant baryon chiral perturbation theory in combination with the so-called $\delta$-expansion power counting scheme \cite{pascal}, which counts $m_\pi$ one order higher in $\delta$ than the delta-nucleon mass splitting. The {\it vector} $N$ to $\Delta\,(1232)$ form factors have already been calculated to leading-one-loop accuracy both in non-relativistic SSE \cite{gellas,tobi} and in the $\delta$-expansion \cite{pv1}.
 
We begin our discussion by considering, in the isospin-symmetric limit, the {\it relativistic} proton to $\Delta^{+}$ matrix element of the isovector axial current
\beq
A_\mu^{(3)}= \bar{q}\, \gamma_\mu \gamma_5 \frac{\tau^3}{2}\,q~,
\eeq
where $q$ is the isospin doublet of the $u$- and $d$-quark fields and the Pauli matrix $\tau^3$ acts in flavor space. On the basis of Lorentz covariance and parity, the relevant amplitude can be expressed in terms of four transition form factors \cite{ls,adler}:
\beqa \label{adler}
{\cal M}^{\rm rel} &=& 
\sqrt{\frac{2}{3}}\;\bar{u}_{\Delta^+}^\lambda(p_\Delta) \left[ \,\frac{C_3(q^2)}{M_N}\, \Big(\epsilon^{(3)}_\lambda \,q\!\!\!/ - q_\lambda\, \epsilon\!\!/^{(3)}\Big) +\frac{C_4(q^2)}{M_N^2}\,\Big(\epsilon^{(3)}_\lambda\, p_\Delta \cdot q - q_\lambda\, \epsilon^{(3)} \cdot p_\Delta \Big) \right. \nonumber \\ && \left. + \,C_5(q^2)\, \epsilon^{(3)}_\lambda + \frac{C_6(q^2)}{M_N^2} \,q_\lambda\, \epsilon^{(3)} \cdot q \right] u_P(p_N)~.
\eeqa
Here $u^\lambda_{\Delta^+}(p_\Delta)$ is a Rarita-Schwinger spinor \cite{RS} and both the proton and the $\Delta^+$ are on-shell. $M_N$ is the nucleon mass, $q^\mu =p_\Delta^\mu-p_N^\mu$ and $\epsilon^{(3)}_\mu$ denotes the polarization vector of the third component of an isotriplet external axial field.

 We evaluate the four $C_{i}\,(q^2,m_\pi)$ to ${\cal O}(\epsilon^3)$ in SSE. This is the lowest order where pion-nucleon and pion-delta loop graphs enter the calculation. Due to the presence of the open $\pi N$ channel, the resulting amplitude develops a non-vanishing imaginary part at the physical pion mass, at variance with the above-mentioned quark model calculations and empirical parameterizations (see also \cite{jorge}).  
From our results for $C_{5}\,(q^2,m_\pi)$  and $C_{6}\,(q^2,m_\pi)$ --- the latter encodes the pion pole contribution --- we derive the off-diagonal Goldberger-Treiman discrepancy to order $\epsilon^3$, cf. \cite{zhu}.
 
This paper is organized as follows. In Section \ref{lag} we briefly review the essentials of the SSE formalism and specify the effective Lagrangian needed for our calculation. Section \ref{fey} is devoted to the discussion of the pertinent Feynman diagrams. We then proceed, in Section \ref{ffs}, to work out the Pauli-reduced transition amplitude and obtain the expressions for the form factors and the off-diagonal Goldberger-Treiman discrepancy. The technical details are discussed in the Appendices.

\section{Effective Lagrangian} \label{lag}

We briefly review the aspects of the SSE formalism which are relevant for our analysis at next-to-next-to-leading order, {\it i.e.} leading-one-loop accuracy. For a more detailed introduction we refer the reader to \cite{HHK}.
In order to specify the effective Lagrangian underlying our calculation, the construction of the third-order axial-$N$-$\Delta$ vertex in SSE is required.

In Ref. \cite{HHK}, the starting point is the {\it relativistic} description of the pion-nucleon-$\Delta\,(1232)$ system  -- at low energies and for small $u$- and $d$-quark masses -- via chiral effective Lagrangians for the isospin doublet Dirac nucleon field $\psi(x)$ and the spin-3/2 isospin-3/2 delta field $\psi_\mu^i(x)$.
The $\Delta\,(1232)$ degrees of freedom are described in terms of a Rarita-Schwinger spinor which transforms according to the representation $D^{\frac{1}{2}} \times D^1$ of the isospin group. The condition $\tau^i \psi^i_\mu=0$ eliminates the isospin-1/2 components. The field $\psi_\mu^i(x)$ as defined in Ref. \cite{HHK} is guaranteed to satisfy all point transformation requirements \cite{point1, point2} by construction. 

Applying Heavy Baryon methods \cite{JM}, both the nucleon and the delta four-momenta are decomposed as
\beq
p^\mu_{N,\, \Delta} = M_0 v^\mu + r^\mu_ {N,\,\Delta}
\eeq
where $M_0$ is the nucleon mass in the $SU(2)$ chiral limit, $v^\mu$ is a time-like unit vector with $v \cdot r \ll M_0$ and $r^\mu$ is a residual soft momentum, {\it i.e.} small as compared to $M_0$ for any $\mu = 0,1,2,3$.
Velocity-dependent fields \cite{georgi} are then defined through the velocity and spin Heavy Baryon projectors \cite{HHK}. 
By integrating out the `small' nucleon and delta components, one derives chiral effective Lagrangians for the `large' fields, $N(x)$ and $T^i_\mu(x)$, defined as
\beqa
N(x)&=&\exp(i M_0\, v \cdot x)\, P_v^+ \,\psi(x) \nonumber \\
T^i_\mu(x) &=&\exp(i M_0\, v \cdot x) \,P_v^+ \, P_{\mu \nu}^{(3)}\, \psi^\nu_i(x)
\eeqa
with
\beqa
P_v^{+}&=&\frac{1 + v\!\!\!/}{2}~, \nonumber \\
P_{\mu \nu}^{(3)}&=& g_{\mu \nu} -\frac{1}{3} \gamma_\mu \gamma_\nu -\frac{1}{3} (v\!\!\!/ \gamma_\mu v_\nu+v_\mu \gamma_\nu v\!\!\!/)~. 
\eeqa
In these Lagrangians, a simultaneous expansion in the number of derivatives, powers of light quark masses and powers of $1/(2M_0)$ is performed. The mass of the $\Delta\,(1232)$ in the $SU(2)$ chiral limit, $M_\Delta^0$, appears through the small parameter $\Delta \equiv M_\Delta^0-M_0 \ll M_0$ which is incorporated in the systematic SSE power counting in $\epsilon$: soft momenta and $\Delta$ count as ${\cal O}(\epsilon)$, $u$- and $d$-quark masses as ${\cal O}(\epsilon^2)$.  

In order to perform the non-relativistic ${\cal O}(\epsilon^3)$ SSE calculation of the axial nucleon to $\Delta\,(1232)$ transition form factors, the required effective Lagrangian 
\beq
{\cal L}={\cal L}_{\pi N}+{\cal L}_{\pi \Delta}+{\cal L}_{\pi N \Delta} + {\cal L}_{\pi \pi}
\eeq
contains both pion-baryon terms up to third order and the pion Lagrangian up to fourth order, due to the contributions from pion pole graphs. All relevant vertices, with the only exception of the third-order axial-$N$-$\Delta$ one, have been constructed already. For completeness and proper definition of the couplings we collect the pertinent Lagrangians.
We use 
\beqa \label{lagrlit}
{\cal L}_{\pi N}^{(1)}&=& \bar{N} (i v \cdot D + g_A \,S\cdot u) N \nonumber \\
{\cal L}_{\pi N}^{(2)}&=&\bar{N} \,c_1 {\rm Tr}(\chi_+)\, N + \dots \nonumber \\
{\cal L}_{\pi N}^{(3)}&=&\bar{N} B_{20}\, [{\rm Tr}(\chi_+) \, i v \cdot D + {\rm h.\,c.}] N+\bar{N} \Delta^2\, B_{30}\, i v \cdot D N + \dots \nonumber \\
{\cal L}_{\pi \Delta}^{(1)}&=&- \bar{T}^\mu _i\,[ i v \cdot D^{i j} - \xi^{i j}\, \Delta + g_1\, S \cdot u\,\delta^{i j}] \,g_{\mu \nu}\,  T^{\nu}_j \nonumber \\
{\cal L}_{\pi \Delta}^{(2)}&=&-\bar{T}^\mu_i \,a_1  {\rm Tr}(\chi_+) \,g_{\mu \nu}\, \delta^{i j}\, T^{\nu}_j + \dots \nonumber\\
{\cal L}_{\pi \Delta}^{(3)}&=&- \bar{T}^\mu_i \,\tilde{B}_{20}\, [(i v \cdot D^{i j} - \xi ^{i j}\,\Delta) {\rm Tr}(\chi_+) + {\rm h.\,c.}] \,g_{\mu \nu}\, T^{\nu}_j \nonumber \\ &&-\bar{T}^{\mu}_i \,\Delta^2\, \tilde{B}_{30} (i v \cdot D^{i j} - \xi^{i j}\, \Delta)\, g_{\mu \nu}\, T^\nu_j + \dots \nonumber 
\eeqa
\beqa
{\cal L}_{\pi N \Delta}^{(1)}&=& c_A \,\bar{T}^\mu_i\, g_{\mu \nu}\, w^\nu_i\,N + {\rm h.\,c.} \nonumber\\
{\cal L}_{\pi N \Delta}^{(2)}&=&\bar{T}^\mu_i \left[ i \,b_3 \,w^i_{\mu \nu}\, v^\nu + i\, b_2\, f^{i\,-}_{\mu \nu}\, v^\nu -\frac{c_A}{M_0}\,i D_\mu^{i j}\, \xi^{j k} \,v \cdot w^k \right] N +\dots \nonumber\\
{\cal L}_{\pi \pi}^{(2)}&=&\frac{F^2}{4} \,{\rm Tr}\!\left[u_\mu u^\mu + \chi_+ \right]  \nonumber \\
{\cal L}_{\pi \pi}^{(4)}&=&\frac{\ell_3}{16} [{\rm Tr}(\chi_+)]^2 +\frac{\ell_4}{16}\left\{2 \,{\rm Tr}(\chi_+) {\rm Tr}(u_\mu u^\mu)+2\, {\rm Tr}(\chi_-^2)-[{\rm Tr}(\chi_-)]^2\right\} +\dots
\eeqa
cf. \cite{HHK, BKM, BFHM, GSS}. The dots denote terms not needed for our purpose. The non-relativistic ${\cal L}_{\pi N \Delta}^{(2)}$ contains the $1/(2M_0)$ `recoil' correction from the first-order Lagrangian in the relativistic formulation \cite{HHK}. $F$, $g_A$, $c_A$ and $g_1$ are, respectively, the pion decay constant, the nucleon axial-vector coupling, the leading $\pi N \Delta$ and $\pi \Delta \Delta$ couplings, all defined in the $SU(2)$ chiral limit; $\xi^{i j}= \delta^{i j}-\tau^i \tau^j/3$ is the isospin-3/2 projection operator. The building blocks for the Lagrangians above are given by
\beqa
u^2(x)&=&\sqrt{1-\frac{\vec{\pi}^{\,2}(x)}{F^2}}+i \,\frac{\vec{\tau} \cdot \vec{\pi}(x)}{F} \nonumber \\
u_\mu&=&i\,\{u^{\dagger},\partial_\mu u\} + u^{\dagger} \,a_\mu\, u+u\, a_\mu\, u^{\dagger}+ \dots\; =
\; \tau^k w^k_\mu \nonumber\\
S_\mu&=&\frac{i}{2}\,\gamma_5\,\sigma_{\mu \nu}\,v^\nu \nonumber \\
D_\mu&=&\partial_\mu+\Gamma_\mu\,, \;\;\;\;\;\;\Gamma_\mu=\frac{1}{2}\,[u^{\dagger},\partial_\mu u]-\frac{i}{2}\, u^{\dagger}\, a_\mu\, u+\frac{i}{2}\, u\, a_\mu u^{\dagger} +\dots \;= \;\tau^k\,\Gamma^k_\mu \nonumber \\
D_\mu^{i j}&=&D_\mu\,\delta^{i j} - 2\, i\,\epsilon^{i j k}\, \Gamma^k_\mu \nonumber\\
w^i_{\alpha \beta}&=&\frac{1}{2}\,{\rm Tr}(\tau^i\,[D_\alpha, u_\beta])\nonumber \\
f^{i\,-}_{\alpha \beta}&=&\frac{1}{2}\,{\rm Tr}\big\{\tau^i\,[ u^{\dagger}(\partial_\alpha\, a_\beta-\partial_\beta\, a_\alpha)\,u + u\,(\partial_\alpha\, a_\beta - \partial_\beta\, a_\alpha)\,u^{\dagger}] \big\}+ \dots \nonumber \\
\chi&=& 2 B\,\hat{m}\,1_{2 \times 2}+\dots\,, \;\;\;\;\;\; \chi_{\pm}=u^{\dagger}\, \chi\, u^{\dagger} \pm u\, \chi^{\dagger}\, u\,,
\eeqa
where $a_\mu(x)=a_\mu^i(x)\,\tau^i/2$ is an isotriplet external axial field. Explicit chiral symmetry breaking due to the light quark masses is encoded in $\chi$. In our analysis we neglect isospin breaking effects and work with $m_u=m_d=\hat{m}$.  

At order $\epsilon^3$, ${\cal L}_{\pi N \Delta}^{(3)}$ receives contributions from the {\it relativistic} third-order Lagrangian, from the second-order one in the form of $1/(2M_0)$ corrections and from the first-order one via $1/(2M_0)^2$ terms.
The ${\cal O}(\epsilon^3)$ $\pi N \Delta$ vertex has been given in Ref. \cite{FM}. Compared to this reference, in the relativistic second- and third-order Lagrangian we have reduced the number of terms by means of integrations by parts and equations of motion (see {\it e.g.} \cite{HHK,FMS}). Furthermore, we have included the ${\cal O}(\epsilon^3)$ counterterms polynomial in $\Delta$ \cite{BFHM} which contribute to our leading-one-loop calculation. 
Constructing the terms which give rise to the ${\cal O}(\epsilon^3)$ axial-$N$-$\Delta$ vertex, we obtain
\beqa
{\cal L}_{\pi N \Delta}^{(3)}\!&=&\!\bar{T}^\mu_i\left\{{\phantom{\frac{M}{N}}}\!\!\!\!\!\!\!f_1\,[D_\mu,\,w^i_{\alpha \beta}]\,v^\alpha v^\beta  + f_4\, w^i_\mu\, {\rm Tr}(\chi_+) + f_5 \,[D_\mu, i\, \chi_-^i] \right. \nonumber \\ &&\left. +\,  f_7\,[f^{i -}_{\mu \beta}, D^{\beta}]+ i D_3\, \frac{\Delta}{M_0} \,w^i_{\mu \beta}\, v^\beta+ i D_2\, \frac{\Delta}{M_0}\, f^{i\,-}_{\mu \beta} \,v^\beta + E_A \,\frac{\Delta^2}{M_0^2}\, w^i_\mu \right. \nonumber \\ && \left. -\,2i \,\frac{b_3}{M_0} \,w^i_{\mu \beta}\, S^\beta\, i S \cdot D - 2i\, \frac{b_3}{M_0}\, i S \cdot D^{ik}\, \xi^{kl}\, w^l_{\mu \beta}\, S^\beta \right. \nonumber \\ &&\left.  -\,2 i\, \frac{b_2}{M_0}\, f^{i\,-}_{\mu \beta}\, S^\beta\, i S \cdot D - 2i\, \frac{b_2}{M_0}\, i S \cdot D^{i k}\, \xi^{k l}\, f^{l\,-}_{\mu \beta}\, S^\beta \right. \nonumber \\ && \left. -\, i\, \frac{b_3}{M_0}\, i D_\mu^{ik}\, \xi^{kl}\, w^l_{\alpha \beta}\, v^\alpha v^\beta - \frac{c_A}{4 M_0^2}\left[\!\!\!\!\phantom{\frac{1}{1}}4\, i S \cdot D^{ik}\, \xi^{kj}\, w^j_\mu\, i S \cdot D \right. \right.  \nonumber \\ && \left. \left. -\, \frac{8+32 z_0}{3}\, i S \cdot D^{ik}\, \xi^{k l}\, i D_\mu^{l m}\, \xi^{m n}\, S \cdot w^n + \frac{32 z_0}{3}\, i D_\mu ^{i k}\, \xi^{k l}\, i S \cdot D^{l m}\, \xi^{m n}\, S \cdot w^n \right. \right. \nonumber \\ && \left. \left. -\, \frac{8 z_0 - 4}{3}\, i D_\mu^{i k}\, \xi^{k l}\, i v \cdot D ^{l m}\, \xi^{m n}\, v \cdot w^n + \frac{8 z_0-16}{3}\, \Delta\, i D_\mu^{i k}\, \xi ^{k l}\, v \cdot w^l \right] \right\} N + \dots
\eeqa
where $\chi^i_{-}=1/2\, {\rm Tr}(\tau^i\,\chi_-)$ and $z_0$ denotes the so-called off-shell parameter appearing in the leading relativistic $\pi N \Delta$ Lagrangian \cite{HHK}.

\section{Feynman diagrams} \label{fey}

\begin{figure}
\begin{center}
  \scalebox{0.75}{\includegraphics*[137,551][533,696]{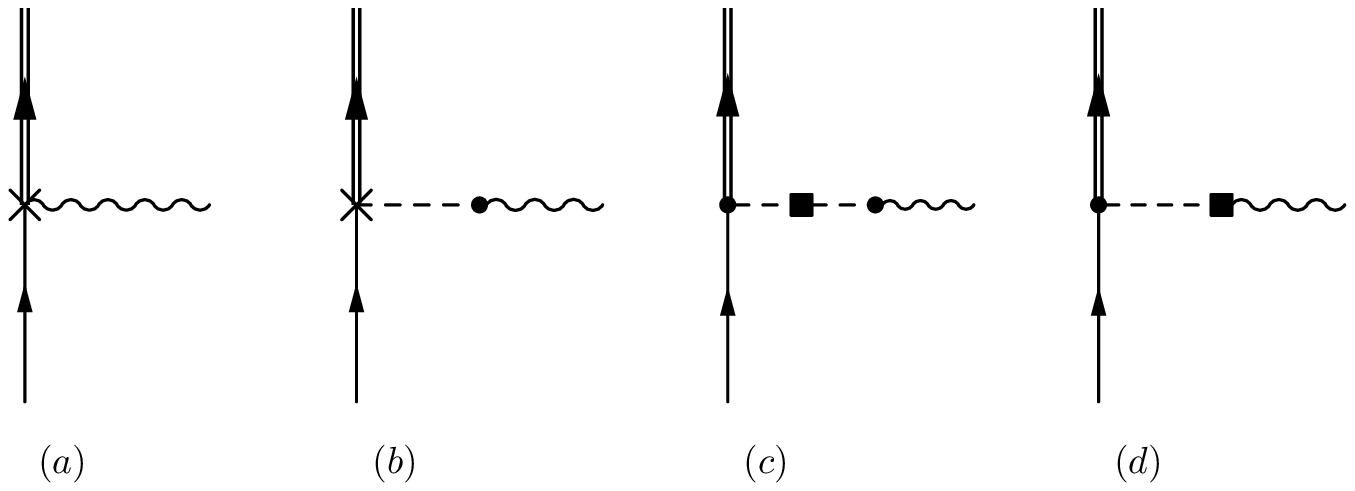}} \\
  \end{center}
     \caption{Tree graphs contributing to the $q^2$ and quark mass dependence of the axial $N$ to $\Delta\,(1232)$ transition form factors to order $\epsilon^3$ in SSE. Dashed, solid and double lines correspond to the pion, the nucleon and the delta, respectively. The wavy line denotes an external axial field. The cross represents vertices from ${\cal L}_{\pi N \Delta}^{(1)}$, ${\cal L}_{\pi N \Delta}^{(2)}$ and ${\cal L}_{\pi N \Delta}^{(3)}$. Contributions from ${\cal L}_{\pi \pi}^{(4)}$ are drawn as squares. The other vertices come from leading order Lagrangians.} \label{tree}
\end{figure}

\begin{figure}
\begin{center}
  \scalebox{0.75}{\includegraphics*[110,207][531,698]{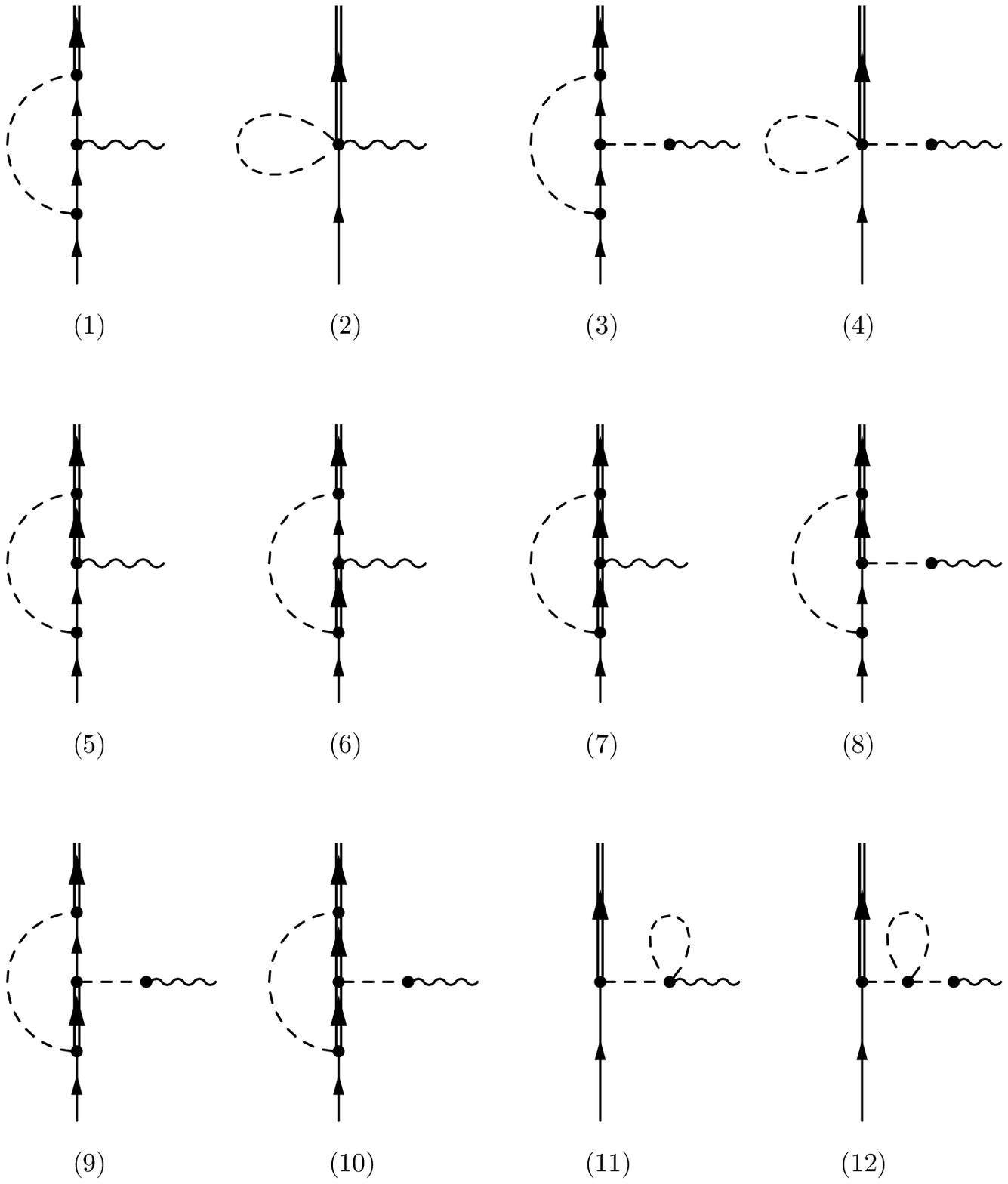}} \\
  \end{center}
     \caption{Loop diagrams relevant for the $q^2$  and quark mass dependence of the axial $N$ to $\Delta\,(1232)$ transition form factors to order $\epsilon^3$ in SSE. All vertices are of leading order. Graphs which vanish are not shown.} \label{loops}
\end{figure}

\begin{figure}
\begin{center}
  \scalebox{0.75}{\includegraphics*[110,553][485,698]{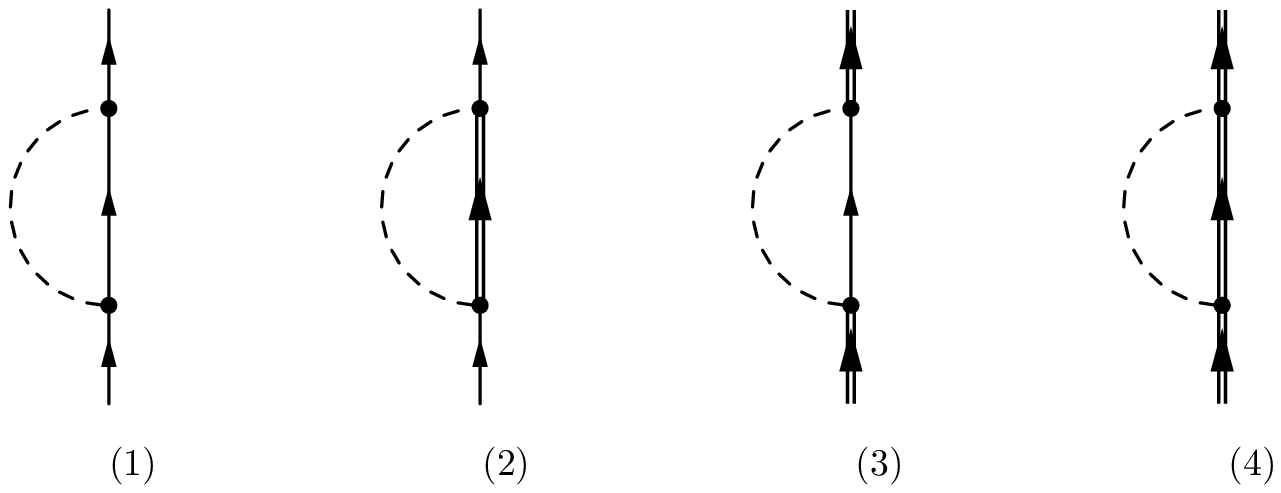}} \\
  \end{center}
     \caption{Loop graphs for nucleon and delta field renormalization to order $\epsilon^3$.} \label{wavef}
\end{figure}

The graphs which are relevant for our calculation are shown in Figs. \ref{tree}, \ref{loops} and \ref{wavef}. The loops in Fig. \ref{wavef} contribute to nucleon and delta field renormalization, namely to $\sqrt{Z_N}$ and $\sqrt{Z_\Delta}$. We do not draw the diagrams which vanish due to the 
light delta constraints:
$v \cdot u_{v,\,\Delta}^i=0$, $S \cdot u_{v,\,\Delta}^i=0$, $\tau^i \,u_{v,\,\Delta}^{\mu,\,i}=0$, where $u_{v,\,\Delta}^{\mu,\,i}(r_\Delta)=P_v^+\,u^{\mu,\,i}_\Delta(p_\Delta)$. 

The loops have been evaluated using dimensional regularization. The results in the rest frame of the $\Delta\,(1232)$ are collected in Appendices \ref{vert} and \ref{amp}. 
Here we point out that:
\begin{itemize}

\item[-] to order $\epsilon^3$, the $q^2$ dependence is only given by counterterms and the pion pole, as in the case of the axial and induced pseudoscalar nucleon form factors \cite{BFHM}.

\item[-] The loop functions in the diagrams (1), (3), (6), (9) of Fig. \ref{loops} and in the one involving $\sqrt{Z_\Delta}$ to order $\epsilon^3$ develop non-vanishing imaginary parts for $m_\pi < \Delta$, as expected since the intermediate $\pi$ and $N$ are there allowed to be simultaneously on-shell.

\item[-] The double pole at $q^2=m_\pi^2$ appearing in diagram (12) of Fig. \ref{loops} has been absorbed (together with higher-order corrections) in the full pion propagator with ${\cal O}(p^4)$ renormalized pion mass $m_\pi^{\rm ren}$, see Appendix \ref{amp} for details. One has \cite{GL}
\beq \label{masspi}
m_{\pi}^{2\,\,{\rm ren}}=m_\pi^2+\frac{m_\pi^4}{F^2}\left(2\,\ell_3^r(\lambda)+\frac{1}{16\pi^2}\ln{\frac{m_\pi}{\lambda}}\right)~,
\eeq
where $m_\pi^2=2 B \hat{m}$ and $\ell_3^r(\lambda)$ denotes the finite part of the coupling $\ell_3$ which depends on the scale $\lambda$ of dimensional regularization. 
\end{itemize}

 \section{The axial $N$ to $\Delta$ transition form factors to order $\epsilon^3$} \label{ffs}
 
We now expand the various terms of the amplitude in Eq. (\ref{adler}) in powers of $1/M_N$ and derive the $N$ to $\Delta\,(1232)$ transition form factors, consistently with the SSE power counting and the accuracy of our calculation, $\it i.e.$ ${\cal O}(1/M_0^2)$.
 For the non-relativistic Pauli reduction of the Dirac and Rarita-Schwinger spinors we follow the notation of Ref. \cite{gellas} (see also \cite{EW}).
In the rest frame of the $\Delta\,(1232)$, with $r_\Delta^\mu=\Delta\,v^\mu$, we find 
\beqa \label{ampnr}
i\,{\cal M}^{\rm non-rel}&=& i\,\sqrt{\frac{2}{3}}\,\bar{u}_{v,\,\Delta^+}^{\mu}(r_\Delta) \left[\epsilon_\mu^{(3)} \left(\frac{C_3}{M_N} \,\Delta_{\rm ph}+\frac{C_4}{M_N}\,\Delta_{\rm ph} + \frac{C_4}{M_N^2} \,\frac{q^2+\Delta^2_{\rm ph}}{2}+ C_5+ {\cal O}\!\left(\frac{1}{M_N^3}\right) \right) \right. \nonumber \\&& \left.+\,q_\mu\,\epsilon^{(3)} \cdot q\, \left(\frac{C_6}{M_N^2}+\frac{C_3}{2 M_N^2}+ {\cal O}\!\left(\frac{1}{M_N^3}\right)\right) \right. \nonumber \\&& \left. +\,q_\mu\, v \cdot \epsilon^{(3)} \left(-\frac{C_3}{M_N}-\frac{C_3}{2 M_N^2}\,\Delta_{\rm ph}-\frac{C_4}{M_N^2}\,\Delta_{\rm ph}-\frac{C_4}{M_N}+ {\cal O}\!\left(\frac{1}{M_N^3}\right) \right) \right.\nonumber \\
&& \left. +\,q_\mu\,[S\cdot \epsilon^{(3)},\,S \cdot q] \left(- \frac{C_3}{M_N^2}+ {\cal O}\!\left(\frac{1}{M_N^3}\right) \right) \right] u_{v,P}(r_N)~.
\eeqa
$M_N$ and $\Delta_{\rm ph}$ denote here the physical nucleon mass and the physical delta-nucleon mass splitting, respectively.

By matching Eq. (\ref{ampnr}) with our tree-level and loop results for the third isospin component of the external axial field and using $M_N(m_\pi)=M_0-4\, c_1 m_\pi^2 + \dots$ and $M_\Delta(m_\pi)=M_\Delta^0-4\,a_1 m_\pi^2+\dots$, we obtain, to the order at which we are working,
\beqa
\frac{C_3}{M_0}&=& -b_3-b_2  \label {c3}\\
\frac{C_4}{M_0^2}&=& 0+ {\cal O}(\epsilon^4) \label{c4} \\
C_5&=&a_1+a_2\,m_\pi^2+a_3\,q^2+ {\rm loop}_{5}(m_\pi) \label{c5} \\
\frac{C_6}{M_0^2}&=&-a_3+\frac{1}{q^2-m_\pi^{2\,{\rm ren}}}\Big(-a_1+\tilde{a}\,m_\pi^2+{\rm loop}_6(m_\pi)\Big) \nonumber \\
&=&\frac{1}{q^2-m_\pi^{2\,{\rm ren}}}\Big(-a_1+a_4\,m_\pi^2-a_3\,q^2 + {\rm loop}_{6}(m_\pi) \Big) \label{c6}
\eeqa
with
\beqa \label{param}
a_1&=& c_A+ b_3\,\Delta -f_1\,\Delta^2+D_3 \frac{\Delta^2}{M_0} +E_A^r(\lambda)\frac{\Delta^2}{M_0^2} -\frac{c_A}{2} \big(B_{30}^r(\lambda) - \tilde{B}_{30}^r(\lambda)\big)\,\Delta^2 \nonumber\\
a_2&=&4 f_4^r(\lambda)-4 \,c_A\,\big(B_{20}^r(\lambda)-\tilde{B}_{20}^r(\lambda)\big) \nonumber \\
a_3&=&-f_7 \nonumber \\
a_4&=&\tilde{a}+a_3\,=\,2\, f_5-4\, f_4^r(\lambda)+4\, c_A\,\big( B_{20}^r(\lambda)-\tilde{B}_{20}^r(\lambda)\big) -f_7 \nonumber \\
{\rm loop}_5(m_\pi)&=&\frac{c_A}{15552 \,\pi^2 F^2}\left\{\frac{1}{\Delta}\,\Big[5\, g_1^2\,(40 \pi\, m_\pi^3 +101 \Delta\, m_\pi^2+ 24 \Delta^3)+1170 \,g_A\, g_1 m_\pi^2\, \Delta \right. \nonumber \\ && \left.-\,12 \Delta\, c_A^2(162\, m_\pi^2-83 \Delta^2)-27\, g_A^2\,(24 \pi\, m_\pi^3+75 \Delta\, m_\pi^2 - 40 \Delta^3)\Big] \nonumber \right. \\ && \left. +\,\frac{72}{\Delta} \sqrt{m_\pi^2-\Delta^2}\, \Big(m_\pi^2\, c_A^2 -28 \Delta^2\, c_A^2+18 g_A^2\, (m_\pi^2 - \Delta^2)\Big) \arccos{\left(-\frac{\Delta}{m_\pi} \right)} \nonumber \right. \\ && \left. -\,\frac{8}{\Delta} \sqrt{m_\pi^2-\Delta^2}\,\Big(9 \,m_\pi^2\, c_A^2+963\, \Delta^2 \,c_A^2+50\, g_1^2\,(m_\pi^2-\Delta^2)\Big) \arccos{\left(\frac{\Delta}{m_\pi} \right)} \nonumber \right. \\ && \left. - \,\Big[3\, m_\pi^2\, (900\, c_A^2-425\, g_1^2-450\, g_1\, g_A +81\, g_A^2 +648)\right. \nonumber \\ && \left.+\,8\Delta^2(-711 \,c_A^2+50\, g_1^2+162\, g_A^2)\Big]\,\ln{\left(\frac{m_\pi}{\lambda}\right)} \right\} \nonumber \\
{\rm loop}_6(m_\pi)&=&-{\rm loop}_5(m_\pi)~.
\eeqa
 
The couplings $b_2,\,b_3,\, f_1,\,f_5$ and $D_3$ possess only a finite part. $B_{20}^r, B_{30}^r, \tilde{B}_{20}^r, \tilde{B}_{30}^r$ are renormalized low-energy constants appearing in the nucleon and delta $Z$-factors. Similarly, ultraviolet divergences in the loops of Fig. \ref{loops} are absorbed by $f_4$ and $E_A$. The renormalized couplings and ${\rm loop}_5(m_\pi)$ depend on the scale of dimensional regularization $\lambda$ in such a way that the right-hand sides of Eqs. (\ref{c5}) and (\ref{c6}) are both $\lambda$-independent.
The expression for ${\rm loop}_5$ is given for the case $m_\pi>\Delta$, relevant for comparison with present lattice QCD results. For its analytic continuation when $m_\pi < \Delta$ we refer to Appendix \ref{amp}.

The low-energy constant $f_7$ represents the `non-pole contribution' to $C_6(q^2)$ and determines the  slope of $C_5(q^2)$. The form factors $C_3$ and $C_4$ acquire $q^2$  and $m_\pi$ dependence only at higher orders in non-relativistic SSE.

An important test of our results is provided by a relation between $C_5$ and $C_6$ required by chiral symmetry. Let us consider the relativistic proton to $\Delta^+$ matrix element of the divergence of the axial current operator $A_\mu^{(3)}(x)$. One has
\beq
\langle \Delta^+(p_\Delta) | \partial^\mu A_\mu^{(3)}(0) | P(p_N) \rangle = i \,\sqrt{\frac{2}{3}}\,\bar{u}^\lambda_{\Delta^+} (p_\Delta)\left[C_5(q^2,m_\pi)+\frac{C_6(q^2,m_\pi)}{M_N^2}\, q^2\right]q_\lambda\, u_P(p_N)~,
\eeq
which, for vanishing quark masses, implies
\beq
C_5(q^2,m_\pi=0)= -\frac{C_6(q^2,m_\pi=0)}{M_0^2} \,q^2~.
\eeq
This constraint is manifestly satisfied by our results.

Let us now define
\beq \label{d}
\frac{D(q^2)}{m_\pi^2-q^2} \equiv C_5(q^2)+\frac{C_6(q^2)}{M_N^2}\,q^2~, 
\eeq
where $m_\pi$ now indicates the physical value of the pion mass. According to Ref. \cite{alex}, 
\beq
D(q^2)= \frac{1}{2 M_N}\, m_\pi^2 F_\pi \,G_{\pi N \Delta}(q^2)~,
\eeq
with $F_\pi=92.4\,{\rm MeV}$ denoting the physical pion decay constant. 
The so-called off-diagonal Goldberger-Treiman discrepancy is then given by
\beq
\Delta_{\rm{ODGT}}=1-\frac{D(q^2=0)}{D(q^2=m_\pi^2)} \simeq m_\pi^2 \,\left. \frac{d}{d q^2} \ln{D(q^2)}\right|_{q^2=m_\pi^2}~,
\eeq
to leading order in $\hat{m}$ \cite{zhu}, see also Ref. \cite{goity}.
From Eqs. (\ref{c5}) and (\ref{c6}) we find, consistently with our ${\cal O}(\epsilon^3)$ accuracy{\footnote{Following instead the notation of Ref. \cite{FM}, we would obtain $\Delta_{\rm ODGT}=-m_\pi^2/ c_A \left[2\,f_5+b_8/(2 M_0)\right]$, where, at variance with \cite{zhu}, all $q^2$ dependent counterterms up to ${\cal O}(\epsilon^3)$ have been taken into account.}},
\beq
\Delta_{\rm{ODGT}}=-\frac{m_\pi^2}{c_A}\,(a_2-a_3+a_4)=-2 f_5\, \frac{m_\pi^2}{c_A}~.
\eeq

In the forthcoming paper \cite{forth} we will compare our Eqs. (\ref{c5}), (\ref{c6}) and (\ref{param}) with the new lattice results for the $Q^2$ and $m_\pi$ dependence of the form factors $C_5$ and $C_6$, where $Q^2=-q^2>0$. While the leading-order parameters $F$, $\Delta$, $c_A$, $g_A$ and $g_1$ are well constrained by hadron phenomenology and/or chiral extrapolations of nucleon observables, the higher-order couplings are not. The determination of the unknown $a_1$, $a_2$, $a_3$, $a_4$ by means of chiral extrapolations will be the subject of the numerical analysis in Ref. \cite{forth}.
  
As already mentioned in the introduction, empirical parameterizations for the $C_i(m_\pi=m_\pi^{\rm phys}, q^2)$ have been used to describe cross section data for neutrino-induced single-pion production off the nucleon. In most of those studies this process is modeled, at intermediate energies, by the $\Delta$-pole mechanism (weak excitation of the delta and its subsequent decay into $N \pi$). Here we stress that, once the relevant effective couplings have been constrained, a stringent comparison between chiral effective field theory and experimental data in the $\Delta\,(1232)$-resonance region is given only by evaluating the {\it full} amplitude of interest, cf. for example \cite{hernandez}. In particular, the SSE calculation of {\it e.g.} the inelastic neutrino scattering process $\nu_\mu\, p \to  \mu^-\,  p\, \pi^+ $ would include, in a systematic fashion, order by order, all background terms required by chiral symmetry  in addition to the $\Delta$-pole mechanism. This study is left for a future publication.

\section{Conclusions} \label{conclusions}

In this paper the $q^2$ and the quark mass expansions of the axial $N$ to $\Delta\,(1232)$ transition form factors have been calculated to leading-one-loop order in the non-relativistic, two-flavor chiral effective field theory known as the `small scale expansion' (SSE). All loop diagrams and counterterm contributions have been systematically analyzed to order $\epsilon^3$.
$C_3$ and $C_4$ turn out to acquire momentum and quark mass dependence only at higher orders in non-relativistic SSE. For the remaining form factors, the $q^2$ dependence to ${\cal O}(\epsilon^3)$ is given by counterterms and the pion pole. For $m_\pi$ smaller than the delta-nucleon mass splitting, both $C_5(q^2, m_\pi)$ and $C_6(q^2, m_\pi)$ have non-vanishing imaginary parts, which are generated by the loop graphs where the intermediate $\pi$ and $N$ are allowed to be simultaneously on-shell. In \cite{forth} the unknown low-energy constants in our expressions for $C_5(q^2, m_\pi)$ and $C_6(q^2, m_\pi)$ will be constrained by performing chiral extrapolations of new lattice data.

We stress that, in order to improve state-of-the-art comparisons between lattice QCD and chiral effective field theory, it is an important task for the future to perform a {\it simultaneous} fit to selected nucleon observables and their uncertainties, with high-statistics lattice results at small momentum transfer and small pion masses. The calculation presented here provides a missing piece of information in view of this global analysis of nucleon observables.

\section{Acknowledgements}

I thank N.~Kaiser and J.~W.~Negele for many stimulating discussions.
Helpful communications with C.~Alexandrou, J.~M.~Camalich, T.~R.~Hemmert, R.~L.~Jaffe, B.~Kubis, C.~Ratti and W.~Weise are also gratefully acknowledged. This work was supported by a Feodor Lynen Fellowship from the Alexander von Humboldt foundation and by U.S. DOE under grant DE-FG02-94ER40818. I thank the MIT Center for Theoretical Physics for hospitality and support.

\clearpage
\appendix
\vspace{2cm} \noindent {\huge\bf Appendix} 
\section{Axial-$N$-$\Delta$ and $\pi$-$N$-$\Delta$  vertices} \label{vert}
We collect the Feynman rules for the axial-nucleon-delta and pion-nucleon-delta vertices up to ${\cal O}(\epsilon^3)$ in the non-relativistic Small Scale Expansion scheme. Here $r^\mu_N$ denotes the residual, soft four-momentum of the incoming nucleon, $r^\mu_\Delta$ is the residual four-momentum of the outgoing delta with isospin index $i$ and $q^\mu = r^\mu_\Delta - r^\mu_N$ is the four-momentum of the incoming pion (or the external axial field) with isospin index $b$. \\

\underline{Axial-N-$\Delta$ vertex}:
\beqa \label{vertax}
i\,\delta^{i b} \!\!\!\! &\Bigg\{\!\!\!\!\! & c_A\, \epsilon_\mu^{(b)}+ b_3\, q_\mu \,v \cdot \epsilon^{(b)} + b_2\, \big(q_\mu \,v \cdot \epsilon^{(b)} - v \cdot q \,\epsilon_\mu^{(b)}\big) -\frac{c_A}{M_0} \,r_\mu^\Delta\, v \cdot \epsilon^{(b)} \nonumber\\  && - \,2\, \frac{b_3}{M_0}\, q_\mu \Big(S \cdot \epsilon^{(b)} \,S \cdot r_N + S \cdot r_\Delta \,S \cdot \epsilon^{(b)}\Big) - \frac{b_3}{M_0} \,r_\mu^\Delta \,v \cdot q \,v \cdot \epsilon^{(b)} \nonumber \\ && - \,\frac{c_A}{M_0^2} \Big[\epsilon_\mu^{(b)} S \cdot r_\Delta\, S \cdot r_N - \frac{2}{3} \,r_\mu^\Delta \,S \cdot r_\Delta \, S \cdot \epsilon^{(b)} + \frac{1-2 z_0}{3}\, r_\mu^\Delta\, v \cdot r_\Delta \, v \cdot \epsilon^{(b)} +\frac{2 z_0-4}{3} \,\Delta\, r_\mu^\Delta\, v \cdot \epsilon^{(b)} \Big] \nonumber \\ && -\,2 \,\frac{b_2}{M_0} \Big( q_\mu\, S \cdot \epsilon^{(b)}\,S \cdot r_N  -  \epsilon_\mu ^{(b)}\, S \cdot q\,S \cdot r_N  +  q_\mu \,S\cdot r_\Delta \,S \cdot \epsilon^{(b)}- \epsilon_\mu^{(b)}\, S\cdot r_\Delta\,S \cdot q \Big) \nonumber \\ && -\,f_1\,q_\mu\, v \cdot \epsilon^{(b)}\,v \cdot q+ 4\,f_4 \, m_\pi^2\, \epsilon_\mu^{(b)} + f_7\,q_\mu\,q \cdot\,\epsilon^{(b)} -f_7\,q^2\,\epsilon_\mu^{(b)}+ D_3\, \frac{\Delta}{M_0}\, q_\mu\, v \cdot \epsilon^{(b)} \nonumber \\ && +\,D_2\, \frac{\Delta}{M_0} \Big(q_\mu\, v \cdot \epsilon^{(b)} - v \cdot q\, \epsilon_\mu^{(b)} \Big) + E_A\, \frac{\Delta^2}{M_0^2}\, \epsilon_\mu^{(b)}\Bigg\}~.
\eeqa

\underline{$\pi$-N-$\Delta$ vertex}:
\beqa \label{vertpion}
\frac{\delta^{i b}}{F} \!\!\!\! &\Bigg\{\!\!\!\!\! & -\, c_A\, q_\mu - b_3 \,q_\mu\, v \cdot q +\frac{c_A}{M_0} \,r_\mu^\Delta\, v \cdot q \nonumber \\ && + \,2 \,\frac{b_3}{M_0}\, q_\mu \Big(S \cdot q \,S \cdot r_N + S \cdot r_\Delta \,S \cdot q \Big) + \frac{b_3}{M_0} \,r_\mu^\Delta\, (v \cdot q)^2 \nonumber \\ && +\,\frac{c_A}{M_0^2} \Big[q_\mu\, S \cdot r_\Delta\, S \cdot r_N - \frac{2}{3}\, r_\mu^\Delta \, S \cdot r_\Delta \, S \cdot q + \frac{1- 2 z_0}{3}\, r_\mu^\Delta\, v \cdot r_\Delta\, v \cdot q + \frac{2 z_0 - 4}{3}\, \Delta \,r_\mu^\Delta \,v \cdot q \Big] \nonumber \\ && +\, f_1\,q_\mu\,(v \cdot q)^2+ 2 \,m_\pi^2\, (f_5- 2 f_4)\, q_\mu - D_3\, \frac{\Delta}{M_0} \,q_\mu\, v \cdot q - E_A \, \frac{\Delta^2}{M_0^2}\Bigg\}~. 
\eeqa

In the rest frame of the $\Delta\,(1232)$, with $v^\mu=(1,0,0,0)$, 
\beqa
r_\mu^\Delta&=&\Delta\,v_\mu~, \nonumber \\
v \cdot q&=&\Delta+\frac{q^2 - \Delta^2}{2 M_0} + {\cal O} \! \left(\frac{1}{M_0^2}\right)
\eeqa
and the diagrams (a) and (b) in Fig. \ref{tree} read
\beqa
{\rm A}_a &=& \!\!\! \bar{u}_{v,\,\Delta}^{\mu,\,i}(r_\Delta)\, i \left\{ \epsilon_\mu^{(i)}\left[c_A\,\sqrt{Z_N}\,\sqrt{Z_\Delta}-b_2\,\Delta +4 m_\pi^2\, f_4-f_7 \,q^2 -D_2\,\frac{\Delta^2}{M_0}+E_A\,\frac{\Delta^2}{M_0^2}\right]\right. \nonumber \\ 
&&\left.+\, q_\mu \,v \cdot \epsilon^{(i)}\left[b_3+b_2+\left(\frac{b_3}{2}+\frac{b_2}{2}+D_3+D_2 \right)\,\frac{\Delta}{M_0} -f_1\,\Delta \right] \right.\nonumber \\
&&+\left. q_\mu\,[S \cdot \epsilon^{(i)},\, S\cdot q] \left( \frac{b_3}{M_0}+ \frac{b_2}{M_0}\right)
+ q_\mu \,\epsilon^{(i)}\cdot q \left(f_7-\frac{b_3}{2 M_0}-\frac{b_2}{2 M_0}\right)\right\}u_v(r_N) \\
{\rm A}_b&=&\!\!\! \bar{u}_{v,\,\Delta}^{\mu,\,i}(r_\Delta) \frac{i}{q^2-m_\pi^2}\, \epsilon^{(i)} \cdot q\,q_\mu \!\left[{\phantom{\frac{1}{1}}}\!\!\!\!\!-c_A\,\sqrt{Z_N}\,\sqrt{Z_\Delta}- b_3\, \Delta+\,2m_\pi^2\, (f_5-2 f_4)\right. \nonumber \\ && \left. +\,f_1\,\Delta^2-D_3\,\frac{\Delta^2}{M_0}-E_A\,\frac{\Delta^2}{M_0^2}\right] u_v(r_N)\,,\label{tree6}
\eeqa 
where we have made use of the light delta constraint $v \cdot u_{v,\,\Delta}^i=0$.

\section{Loop graphs} \label{amp}

We collect here the loop integrals relevant for our analysis. They have been evaluated using dimensional regularization. Following the notation of Ref. \cite{BKM}:
\beqa \label{loopfunc}
\Delta_\pi &\equiv& \frac{1}{i} \int\, \frac{d^d l}{(2 \pi)^d}\, \frac{1}{m_\pi^2 - l^2 -i \epsilon} = 2 m_\pi^2\left(L + \frac{1}{16 \pi^2} \ln \frac{m_\pi}{\lambda} \right)+ {\cal O}(d-4)~, \nonumber \\
J_0(\omega) & \equiv& \frac{1}{i} \int \frac{d^d l}{(2 \pi)^d} \,\frac{1}{(v \cdot l - \omega - i \epsilon)\, (m_\pi^2 - l^2 - i \epsilon)} \nonumber \\ & =& -\,4 L\, \omega + \frac{\omega}{8 \pi^2} \left(1-2 \ln \frac{m_\pi}{\lambda}\right)\nonumber \\ 
&&- \frac{1}{4 \pi^2} \times
\left\{\begin{array}{l}
\sqrt{m_\pi^2 - \omega^2}\, \arccos\left(-\frac{\omega}{m_\pi}\right) + {\cal O}(d-4)~,\;\;\;\;\;\;\;\;\;\;\;\;\;\;\;\;\;\;\;\;\;\;\;\;\,\,\,\omega^2 \leq m_\pi^2 \nonumber \\ \nonumber \\
-\sqrt{\omega^2-m_\pi^2}\, \ln \left(-\frac{\omega}{m_\pi}+\sqrt{\frac{\omega^2}{m_\pi^2}-1}\,\right)+ {\cal O}(d-4) ~,\;\;\;\;\;\;\;\;\;\;\,\omega <- m_\pi    \nonumber \\ \nonumber\\
\sqrt{\omega^2-m_\pi^2}\,\left[\ln \left(\frac{\omega}{m_\pi}+\sqrt{\frac{\omega^2}{m_\pi^2}-1}\,\right) - i\,\pi\right]+ {\cal O}(d-4) ~,\;\;\;\;\; \omega > m_\pi
\end{array} \right. 
\eeqa
\beqa
g_{\mu \nu}\,J_2(\omega) &\!\!+&\!\! v_\mu v_\nu \,J_3(\omega) \equiv \frac{1}{i} \int \frac{d^d l}{(2 \pi)^d} \,\frac{l_\mu l_\nu}{(v \cdot l - \omega - i \epsilon) (m_\pi^2 - l^2 - i \epsilon)}~,\nonumber \\
J_2(\omega)& =&\frac{1}{d-1} \Big[(m_\pi^2 - \omega^2) J_0(\omega) - \omega \Delta_\pi\Big]~,
\eeqa
where $d$ is the space-time dimension. Any ultraviolet divergence appearing in the limit $d \to 4$ is subsumed in 
\beq
L = \frac{\lambda^{d-4}}{16 \pi^2} \left[\frac{1}{d-4} - \frac{1}{2} \Big(\!\ln{(4 \pi)} + \Gamma'(1)+1\Big) \right]~. 
\eeq

According to the notation of Appendix \ref{vert} for momenta and isospin indices, we give the expressions for the loop graphs in Fig. \ref{loops} evaluated in the $\Delta\,(1232)$ rest frame, with $v^\mu = (1,0,0,0)$ and $r^\mu_\Delta=\Delta \, v^\mu$: 
\beqa
{\rm A}_1 &=&\bar{u}_{v,\,\Delta}^{\mu,\, i} (r_\Delta)\,i\, \frac{ c_A g_A^2}{F^2}\, \epsilon_{\mu}^{(i)}\,\left[\frac{1}{\Delta} J_2(\omega=0) - \frac{1}{\Delta} J_2(\omega= \Delta)\right]\,u_v(r_N) \nonumber \\
{\rm A}_2 &=&\bar{u}^{\mu, \,i}_{v,\,\Delta}(r_\Delta) \left(- i\,\frac{c_A}{F^2} \right)\, \epsilon_{\mu}^{(i)} \,\Delta_\pi \,u_v(r_N)\nonumber \\
{\rm A}_3 &=&\bar{u}^{\mu,\,i} _{v,\,\Delta}(r_\Delta)\left(-i\, \frac{c_A g_A^2}{F^2}\right) \frac{\epsilon^{(i)}\cdot q}{q^2 - m_\pi^2}\, q_\mu\, \left[\frac{1}{\Delta} J_2(\omega=0) - \frac{1}{\Delta} J_2(\omega= \Delta)\right] u_v(r_N)\nonumber \\
{\rm A}_4 &=&\bar{u}^{\mu,\,i}_{v,\,\Delta}(r_\Delta) \left(-i\, \frac{c_A}{2 F^2} \right) \frac{\epsilon^{(i)}\cdot q}{q^2- m_\pi^2}\, q_\mu \, \Delta_\pi \,u_v(r_N)\nonumber \\
{\rm A}_5 &=&\bar{u}^{\mu,\,i}_{v,\,\Delta}(r_\Delta) \left(-i\,\frac{5 g_1 c_A g_A}{3 F^2}\right)  \epsilon_{\mu}^{(i)} \left[\frac{d^2-2d-3}{4(1-d)} \right]\, \left.\left({\phantom{\frac{\partial}{\partial \omega}}}  \!\!\!\!\!\!\!\!\!\!-J_2'(\omega)\right|_{\omega=0}\, \right) u_v(r_N)\nonumber \\
{\rm A}_6 &=&\bar{u}^{\mu,\,i}_{v,\,\Delta}(r_\Delta)\,i\, \frac{c_A^3}{3 F^2}\, \epsilon_{\mu}^{(i)}\, \frac{d-3}{d-1} \left[\frac{1}{2 \Delta} J_2(\omega=-\Delta) - \frac{1}{2 \Delta} J_2(\omega=\Delta)\right] u_v(r_N)\nonumber \\
{\rm A}_7 &=& \bar{u}^{\mu,\,i} _{v,\Delta}(r_\Delta)\,i\,\frac{10 g_1^2 c_A}{9 F^2}\, \epsilon_{\mu}^{(i)} \left[\frac{d^2-2d-3}{2(d-1)^2} \right] \left[\frac{1}{\Delta} J_2(\omega=-\Delta)-\frac{1}{\Delta}J_2(\omega=0) \right] u_v(r_N) \nonumber \\
{\rm A}_8 &=&\bar{u}^{\mu,\,i}_{v,\,\Delta}(r_\Delta)\,i\,\frac{5 g_1 c_A g_A}{3 F^2} \,\frac{\epsilon^{(i)} \cdot q}{q^2-m_\pi^2}\,q_\mu\left[\frac{d^2-2d-3}{4(1-d)}\right]\, \left.\left({\phantom{\frac{\partial}{\partial \omega}}}  \!\!\!\!\!\!\!\!\!\!-J_2'(\omega)\right|_{\omega=0}\, \right) u_v(r_N) \nonumber \\
{\rm A}_9 &=&\bar{u}^{\mu,\,i}_{v,\,\Delta}(r_\Delta) \left( - i \frac{c_A^3}{3 F^2} \right) \frac{\epsilon^{(i)} \cdot q}{q^2-m_\pi^2}\,q_\mu\, \frac{d-3}{d-1}\left[\frac{1}{2 \Delta} J_2(\omega=-\Delta) - \frac{1}{2 \Delta} J_2(\omega=\Delta)\right] u_v(r_N) \nonumber \\
{\rm A}_{10} &=&\bar{u}^{\mu,\,i}_{v,\,\Delta}(r_\Delta)\left(-i \frac{10 g_1^2 c_A}{9 F^2} \right) \frac{\epsilon^{(i)} \cdot q}{q^2-m_\pi^2}\,q_\mu \left[\frac{d^2-2d-3}{2(d-1)^2}\right]\left[\frac{1}{\Delta} J_2(\omega=-\Delta)-\frac{1}{\Delta}J_2(\omega=0) \right] u_v(r_N)\nonumber \\
{\rm A}_{11} &=&\bar{u}^{\mu,\,i}_{v,\,\Delta}(r_\Delta)\,i\,\frac{c_A}{2 F^2}\, \frac{\epsilon^{(i)} \cdot q}{q^2-m_\pi^2}\,q_\mu\,\Delta_\pi\,u_v(r_N)\nonumber \\
{\rm A}_{12} &=&\bar{u}^{\mu,\,i} _{v,\,\Delta}(r_\Delta)\,i\,\frac{c_A}{2 F^2}\,\frac{\epsilon^{(i)} \cdot q}{q^2 - m_\pi^2}\,q_\mu\,\frac{2q^2 - 3m_\pi^2}{q^2-m_\pi^2}\,\Delta_\pi\,u_v(r_N)~.
\eeqa
The diagrams where the intermediate $\pi$ and $N$ are allowed to be simultaneously on-shell give rise to an absorptive piece in the amplitude for $m_\pi < \Delta$.

The factor $1/(q^2-m_\pi^2)^2$ in ${\rm A}_{12}$ is absorbed by using the full pion propagator. The sum of diagram (12) in Fig. \ref{loops}, diagram (c) and the leading term of diagram (b) in Fig. \ref{tree} equals indeed, up to higher-order corrections,
\beq
\bar{u}^{\mu,\,i}_{v,\,\Delta}(r_\Delta) \,(-i\,c_A)\,\frac{\epsilon^{(i)} \cdot q}{q^2-m_\pi^{2\,{\rm ren}}} \,q_\mu\,Z_\pi\,u_v(r_N)~,
\eeq
where
\beq
m_\pi^{2\,{\rm ren}}=m_\pi^2+\frac{\Delta_\pi}{2 F^2}\, m_\pi^2+ 2\, \ell_3\,\frac{m_\pi^4}{F^2}~, 
\eeq
\beq
Z_\pi=1-\frac{\Delta_\pi}{F^2} - 2\, \frac{\ell_4}{F^2}\, m_\pi^2~.
\eeq
In Eq. (\ref{masspi}), $\ell_3^r(\lambda)=\ell_3+ L(\lambda)/2$.
Note that the low-energy constant $\ell_4$ enters at tree-level also in the ${\cal O}(\epsilon^3)$ pion-pole graph with axial-pion coupling from ${\cal L}_{\pi \pi}^{(4)}$, diagram (d) in Fig.\ref{tree}. 

The loops in Fig. \ref{wavef} determine nucleon and delta $Z$-factors to order $\epsilon^3$: 
\beqa
\sqrt{Z_N}&=&1+\frac{1}{2}\,\left.{\phantom {\frac{1}{1}}}\!\!\!\!\!\Sigma_N'(\omega)\right|_{\omega=0} \\
\sqrt{Z_\Delta}&=&1+\frac{1}{2}\,\left.{\phantom {\frac{1}{1}}}\!\!\!\!\!\Sigma_\Delta'(\omega)\right|_{\omega=\Delta}~, \label{Zdelta}
\eeqa
where the relevant contributions to the nucleon and delta self-energies are given by
\beqa
\Sigma_{N,\,1}^{\rm loop}(\omega)&=&\frac{3 g_A^2}{4 F^2}\, \Big[(m_\pi^2-\omega^2) J_0(\omega)- \omega \Delta_\pi \Big] \\
\Sigma_{N,\,2}^{\rm loop}(\omega)&=&\frac{2 c_A^2}{F^2} \,(d-2)\, J_2(\omega-\Delta)\\
\Sigma_{\Delta,\,3}^{\rm loop}(\omega)&=&-\frac{c_A^2}{F^2}\, J_2(\omega)\\
\Sigma_{\Delta,\,4}^{\rm loop}(\omega)&=&\frac{5 g_1^2}{3 F^2}\,\frac{d^2-2d-3}{4(1-d)} \,J_2(\omega-\Delta)~.
\eeqa
The pertinent counterterms are shown in Eq. (\ref{lagrlit}). As expected, $\Sigma_{\Delta,\,3}$ and $Z_\Delta$ have non-vanishing imaginary parts for $m_\pi < \Delta$. 


\end{document}